\documentclass[aps,prb,reprint, amsmath,amssymb,superscriptaddress]{revtex4-2}
\usepackage{graphicx}

\usepackage{graphicx}   % Include figure files
\usepackage{dcolumn}    % Align table columns on decimal point
\usepackage{bm}         % bold math
\usepackage{amsmath}    % AMS math
\usepackage{amssymb}    % AMS symbols
\usepackage{mathtools}

\newcommand{\Gk}{$G_\mathbf{k}(\tau)$}
\newcommand{\gk}{G_\mathbf{k}(\tau)}

\newcommand{\Ak}{$A_\mathbf{k}(\omega)$}

\begin{document}

\title{Single-Particle Dispersion and Density of States \\ of the Half-Filled 2D Hubbard Model}

\author{Gabe Schumm}
\email{gschumm@bu.edu}
\affiliation{Department of Physics, Boston University, 590 Commonwealth Avenue, Boston, Massachusetts 02215, USA}

\author{Shiwei Zhang}
\email{szhang@flatironinstitute.org}
\affiliation{Center for Computational Quantum Physics, Flatiron Institute, 162 5th Avenue, New York, New York, USA}

\author{Anders W. Sandvik}
\email{sandvik@bu.edu}
\affiliation{Department of Physics, Boston University, 590 Commonwealth Avenue, Boston, Massachusetts 02215, USA}
\affiliation{School of Physical and Mathematical Sciences, Nanyang Technological University, Singapore}

\date{\today}

\begin{abstract}
Implementing an improved method for analytic continuation and working with imaginary-time correlation functions computed
using quantum Monte Carlo simulations, we resolve the single-particle dispersion relation and the density of states (DOS) of the
two-dimensional Hubbard model at half-filling. At intermediate interactions of $U/t = 4,6$, we find quadratic dispersion around
the gap minimum at wave-vectors $\mathbf{k} = (\pm \pi/2, \pm \pi/2)$ (the $\Sigma$ points). We find saddle points at
$\mathbf{k} = (\pm \pi,0),(0,\pm \pi)$ (the X points) where the dispersion is approximately quartic,
leading to a sharp DOS maximum above the almost flat ledge arising from the states close to $\Sigma$. The fraction of quasiparticle states
within the ledge is $n_{\rm ledge} \approx 0.15$. Upon doping away from half-filling, within the rigid-band approximation, these results support
Fermi pockets around the $\Sigma$ points, with states around the X points becoming filled only at doping fractions $x \ge n_{\rm ledge}$. The high
density of states away from the $\Sigma$ gap edge may be an important clue for a finite minimum doping level for superconductivity and other
instabilities of doped Mott insulators.
\end{abstract}

\maketitle

\section{Introduction}
The Hubbard model \cite{hubbard_63,kanamori_63,gutzwiller_63} serves as the simplest and most essential model for the physics
of correlated electrons, with the two-dimensional (2D) square-lattice case being of particular interest in the context of the unresolved puzzle of
superconductivity in the cuprates \cite{lee_06,fradkin_15,proust_19,zhou_21}. We focus here on half filling, where auxiliary-field quantum Monte Carlo
(AFQMC) simulations can access the ground state of relatively large systems.
Significant progress has been made on static observables  \cite{hirsch_85, hirsch_89, white_89_2, white_91_2, moreo_90, moreo_91, varney_09, qin_16}, but accessing dynamics, e.g., the important single-particle
spectral function, is much more challenging \cite{qin_22}. While AFQMC simulations \cite{foulkes_01,zhang_13} can be employed to
calculate the wave-vector (${\bf k}$) resolved imaginary-time Green's function \cite{hirsch_85, white_89, moreo_90,scalapino_92,bulut_94,vitali_16}, the corresponding real frequency spectral function $A_{\bf k}(\omega)$ has been difficult to extract because of the ill-posed analytic
continuation problem. Though some key aspects of the dispersion relation have been obtained \cite{bulut_94,preuss_95,assaad_98,assaad_99},
significant uncertainties remain, and a precise characterization is still lacking. Within the rigid band approximation (RBA) \cite{stern_67, eder_94, eder_96}, detailed knowledge of $A_{\bf k}(\omega)$ is required to understand the manner in which the quasiparticle states of the half-filled system are occupied
upon doping and how important scattering channels emerge. This is essential for a coherent understanding of the model and its connection to
high-temperature superconductivity and other phenomena in doped Mott insulators.

The specific technical challenge of computing $A_{\bf k}(\omega)$ is that narrow quasiparticle peaks and associated sharp features in the
density of states (DOS) cannot be reproduced by analytic continuation of AFQMC data with the conventional maximum-entropy method (MEM)
\cite{silver_90_1,silver_90_2,gubernatis_91,gubernatis_96,bergeron_16} or the related stochastic analytic continuation (SAC) (or average spectrum) method
\cite{white_91,sandvik_98,beach_04,vafayi_07,reichman_09,olav_08,fuchs_10,qin_17,ghanem_20,koch_20,ghanem_20,ghanem_23}. However, recent
extensions of SAC have shown that this shortcoming can in many cases be overcome by appropriately constraining the sampling space to
favor sharp features, e.g., peaks and edges that often appear in ground-state spectral functions of quantum many-body
systems \cite{sandvik_16,sandvik_17,shao_23,schumm_24,yang_24}.

In this work, we implement constrained SAC to extract the dispersion relation and the DOS of the half-filled Hubbard model with sufficient
precision to uncover features of key significance for understanding the Mott insulating state. Moreover, within the RBA, the spectral function also
dictates the emergence and initial evolution of the Fermi sea upon doping. Focusing on
intermediate values $U/t = 4,6$ of the Hubbard repulsion, we find a clear separation between the lowest single-particle energy $\omega_\Sigma$
at the four equivalent wave-vectors $\mathbf{k} = (\pm \pi/2, \pm \pi/2)$ (the $\Sigma$ points) and almost dispersionless excitations at
$\omega \approx \omega_{\rm X}$ close to $\mathbf{k} = (\pm \pi,0),(0,\pm \pi)$ (the X points). With the dispersion being quadratic for $\mathbf{k}$
close to the $\Sigma$ points, there is a ledge of almost constant DOS in the range $\omega \in [\omega_\Sigma,\omega_{\rm X})$, followed by a sharp
edge with inverse square-root divergence at $\omega = \omega_{\rm X}$ from the almost flat (nearly quartic) dispersion around the X points. We are able to
determine the DOS to high precision and extract the total fraction $n_{\rm ledge}$ of quasiparticle states in the ledge, with $n_{\rm ledge}= 0.13$
and $n_{\rm ledge} = 0.20$ for $U/t=4$ and $U/t=6$, respectively.

Interpreting our results within the RBA (the validity of which we will also discuss, though our calculations do not address it directly),
the density of quasiparticle states below $\omega_{\rm X}$ controls
the critical doping fraction $x=x_c=n_{\rm ledge}$ for the initial occupation of states around the X points. With the large DOS at
$\omega \approx \omega_{\rm X}$ and the important $(\pi,\pi)$ magnon scattering processes connecting X points, this doping fraction may
signify an instability, with superconductivity and charge-density-wave (CDW) ``stripes'' being natural candidates \cite{xu_24}. 

\section{Model and methods}
In standard notation, the Hamiltonian for the half-filled Hubbard model is
\begin{equation}
  H =-t \sum_{\langle i, j \rangle, \sigma} \hat{c}_{i, \sigma}^{\dagger} \hat{c}_{j, \sigma}+
  U \sum_i(\hat{n}_{i, \uparrow}-\hbox{$\frac{1}{2}$})(\hat{n}_{i, \downarrow}-\hbox{$\frac{1}{2}$}),
\end{equation}
here on the periodic square lattice with $N=L^2$ sites. We set $t=1$ and use the AFQMC method to compute the imaginary-time dependent Green's function,
\begin{equation}\label{Gtau}
G_\mathbf{k}(\tau) =  \langle T_{\tau} c_\mathbf{k}(\tau)c_\mathbf{k}^\dagger(0)\rangle,
\end{equation}
at temperatures $T$ low enough to converge to the ground state. We have confirmed that $\beta=1/T=2L$ is sufficient for all practical purposes
for the repulsion strengths and system sizes $L \le 20$ used here. In the AFQMC simulations, we use a time slice $\Delta\tau =0.1$ and compute
$G_\mathbf{k}(\tau)$ on this grid. The discretization error scales as $\Delta_\tau^2$ and is insignificant, which we have confirmed using spot-checks of $\Delta\tau <0.1$ for some of the smaller lattices. For the cases we study here (i.e., half-filling), AFQMC is sign-free \cite{loh_90} and
numerically exact apart from the discretization error.

The single-particle spectral function \Ak\ is related to \Gk\ via the inverse transform
\begin{equation}\label{transform}
G_\mathbf{k}(\tau) = \int_{-\infty}^{\infty} d\omega \, \frac{e^{-\tau \omega}}{1 + e^{-\beta \omega}} \,A_\mathbf{k}(\omega),
\end{equation}
which we invert for \Ak\ using the SAC method. The DOS, $D(\omega)$, is the average of \Ak\ over $\mathbf{k}$, which in imaginary time corresponds to
\begin{equation}
\frac{1}{N}\sum_k \gk =  \frac{1}{N}\sum_{\mathbf{k}, \mathbf{r}} G_\mathbf{r}(\tau)e^{i \mathbf{k}\cdot \mathbf{r}} \equiv G_{\mathrm{loc}}(\tau),
\end{equation}
where $G_{\mathrm{loc}}(\tau) = G_{{\mathbf{r} = \mathbf{0}}}(\tau)$ is the local Green's function. Particle-hole symmetry at half-filling implies
$D(-\omega) = D(\omega)$, which allows us to preform the analytic continuation of $G_{\mathrm{loc}}(\tau)$ on only the positive frequency axis,
by implementing a modified kernel in Eq.~\eqref{transform}:
\begin{equation}\label{phs_transform}
G_\mathrm{loc}(\tau) = \int_{0}^{\infty} d\omega \, \left(\frac{e^{-\tau \omega}}{1 + e^{-\beta \omega}} + \frac{e^{\tau \omega}}{1 + e^{\beta \omega}}
\right) \,D(\omega). 
\end{equation}

\begin{figure}[t]
\includegraphics[width=0.95\columnwidth]{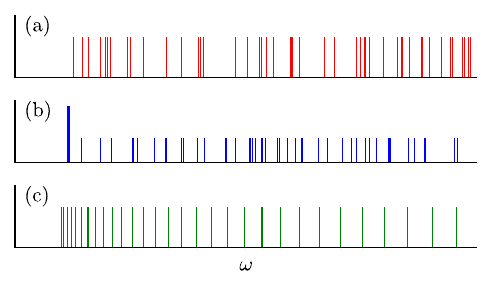}
\caption{Schematic depictions of the $\omega>0$ part of SAC sampling spaces, with the spectrum parametrized as a sum of $\delta$-function
with (a) unconstrained frequencies and fixed amplitudes, (b) a macroscopic $\delta$-function (quasi particle) at $\omega_0$, which also acts
as a lower bound for the other (continuum) contributions, and (c) constrained so that the distance between adjacent $\delta$-functions is monotonically
increasing. This constraint produces an average spectrum with a sharp (divergent for $\omega \to \omega_0^+$ when $N_\omega \to \infty$) edge
followed by an monotonically decaying continuum.}
\label{params}
\end{figure}

We here only briefly summarize our implementation of the SAC method for the problem at hand and refer to Ref.~\onlinecite{shao_23}
for technical details. In the basic formulation, the spectrum \Ak\ is parametrized by a large number $N_\omega$ of $\delta$-functions carrying
weights $a_i$ at energies $\omega_i$, which can take continuous values, as illustrated in Fig.~\ref{params}(a) in the case of uniform amplitudes. Typically
$N_\omega$ is of order $10^3$ or larger. The energies and, optionally, the amplitudes are importance-sampled according to a Boltzmann-like probability
distribution
\begin{equation}
P(A_\textbf{k}) \propto {\rm exp}[-\chi^2(A_\textbf{k})/2\Theta],
\end{equation}
and the spectrum is accumulated as a histogram.
The goodness-of-fit $\chi^2$ is calculated with respect to the QMC-generated Green's function $G_\textbf{k}(\tau_i)$, $i=1,\ldots,N_\tau$,
and involves the full covariance matrix to account for correlatated fluctuations in imaginary time \cite{gubernatis_96}. The fictitious
temperature $\Theta$ in the probability distribution is adapted according to a simple criterion, motivated by properties of the $\chi^2$-distribution,
to properly balance goodness-of-fit and entropy, thus avoiding overfitting while guaranteeing a $\langle \chi^2\rangle$ value representing a
good statistical fit.

Like the MEM, with typical data quality, unrestricted SAC can only produce smooth spectral features. The key insight allowing
for the resolution of sharp features is that various constraints can be imposed on the amplitudes and locations of the $\delta$-functions. The
associated changes in entropic pressures under constraints, or with different parametrizations (e.g., with or without updates of the amplitudes)
impact the exact form of the average spectral density, along with the information contained in $G_\textbf{k}(\tau)$.
As an example, Fig.~\ref{params}(b) depicts a sampling
space constrained such that a ``macroscopic'' $\delta$-function of relative weight $a_0$ at position $\omega_0$ acts as a hard lower bound to a
continuum parametrized just as in the unconstrained case. Here, the $N_\omega$ ``microscopic'' equal amplitude $\delta$-functions each have weight
$(1-a_0) /N_\omega$, where $a_0$ is fixed (and later optimized) but the edge location $\omega_0$ is sampled. To determine the optimal quasiparticle
weight, we scan over $a_0$, as will be detailed below. Unless the optimal $a_0$ is very small, $\omega_0$ fluctuates
very little once it has equilibrated to its optimal position.

In the particle-hole symmetric half-filled Hubbard model, there will be two quasiparticle peaks, at $\omega_0 = \pm |\omega_\mathbf{k}|$,
which define the dispersion relations for injected holes and particles. In this case, the optimized $a_0$ represents the sum of the
two weights and the relative distribution between positive and negative part is sampled. This constrained parametrization
is suitable under the assumption of the true quasiparticle peak being very narrow, which can be expected here at least close to the
minimum $|\omega_\mathbf{k}|$.

Figure \ref{params}(c) shows a different type of constrained parameterization, where it is imposed that the spacing between adjacent
$\delta$-function increases monotonically with $\omega$. As a better alternative to collecting spectral weight in a histogram, the mean spectral
density can in this case be defined for $i=1,\ldots,N_\omega$ as
\begin{equation}
S(\omega_{i+1/2})=\frac{1}{2}\frac{a_{i}+a_{i+1}}{\langle \omega_{i+1}-\omega_{i}\rangle},
\end{equation}
where $\omega_{i+1/2}=\langle \omega_{i}+\omega_{i+1}\rangle/2$ defines the self-generated grid upon which the spectral function is evaluated.
With uniform $a_i$, the monotonicity constraint implies an entropic pressure leading to a singularity (strictly for $N_\omega \to \infty$);
$A(\omega \to \omega_0) \propto (\omega-\omega_0)^{-1/2}$ \cite{shao_23}, where we have defined the edge location $\omega_0\equiv \langle \omega_1\rangle$. Away from the edge, the spectrum adapts according to $G_\mathbf{k}(\tau)$. We will use an extended form of this parametrization for the DOS, after
discussing our results for the dispersion relation.

\begin{figure}[t]
\includegraphics[width=0.98\columnwidth]{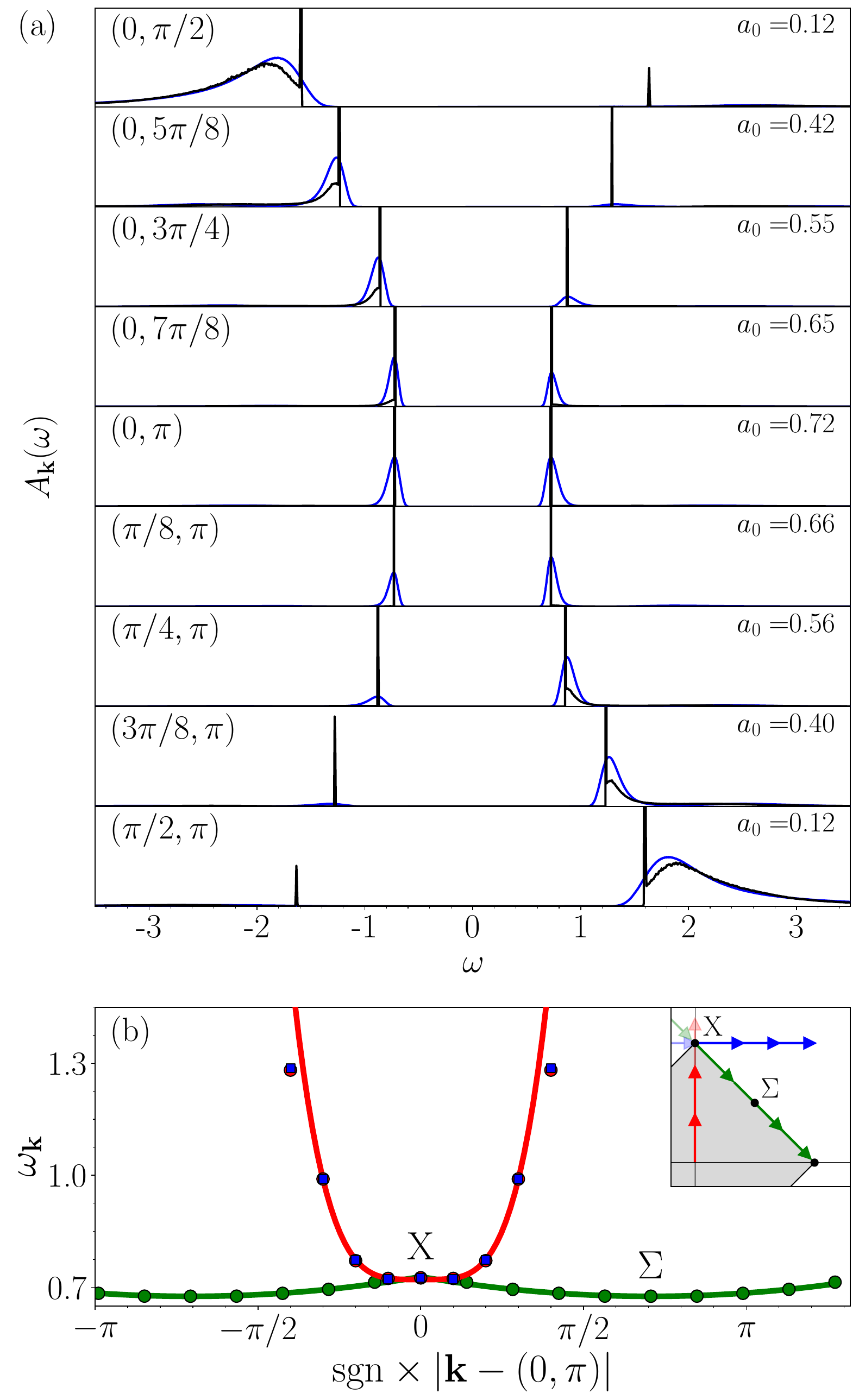}
\caption{(a) \Ak\ for a series of $\mathbf{k}$ points in an $L=16$ system with $U/t=4$, comparing results of  unconstrained (blue) and $\delta$-edge
constrained (black) parameterizations. The optimal macroscopic quasiparticle weights $a_0$ are indicated in each case. The spectra have been rescaled so that the maximum values of the unconstrained spectra are normalized to unity.~(b) The dispersion relation obtained from the $\delta$-edge locations, where
the different colors correspond to the $\mathbf{k}$-space cuts depicted in the inset (sig on $x$-axis indicates coming in $(-)$ or going out $(+)$ of the reference $\mathbf{k}$-point, $(0, \pi)$). The red and blue symbols coincide. The green and red curves
are fits of the form $~\omega_{\Sigma} + aq^2$ and $~\omega_{\rm X} + bq^4$, respectively, with $q$ being the distance to the respective
reference points.}
\label{Ak}
\end{figure}

\section{Single-Particle Spectral Function}

We first examine \Ak\ with $\textbf{k}$ along lines of high symmetry in the Brillouin zone (BZ).
Results for the $L=16$ system at $U/t=4$ obtained with both unconstrained sampling and the $\delta$-edge constraint are presented in Fig.~\ref{Ak}(a).
The spectra from unconstrained SAC are qualitatively very similar to previous results obtained with the MEM \cite{bulut_94,preuss_95,assaad_98,assaad_99}. The spectra generated using the  $\delta$-edge parameterization, on the other hand, differ significantly. For each momentum, we perform a scan over the combined macroscopic $\delta$-function weight $a_0$ and identify an optimal value, as described in Sec.~\ref{a0_scan_Ak} below. These optimal weights are displayed in the top right of each panel. At the lowest energies, the peaks are sufficiently narrow for their centers to coincide with the location of the $\delta$-edge of the constrained spectrum. Correspondingly, the continuum weights of these constrained spectra are relatively small. However, at the higher energies the peaks are too broad to provide a reliable dispersion relation, due to the dominant continuum beyond the quasiparticle peak. 

\subsection{Optimization of $a_0$ for the Single-Particle Spectral Function}\label{a0_scan_Ak}

\begin{figure}[t]
\centering
\includegraphics[width=0.95\columnwidth]{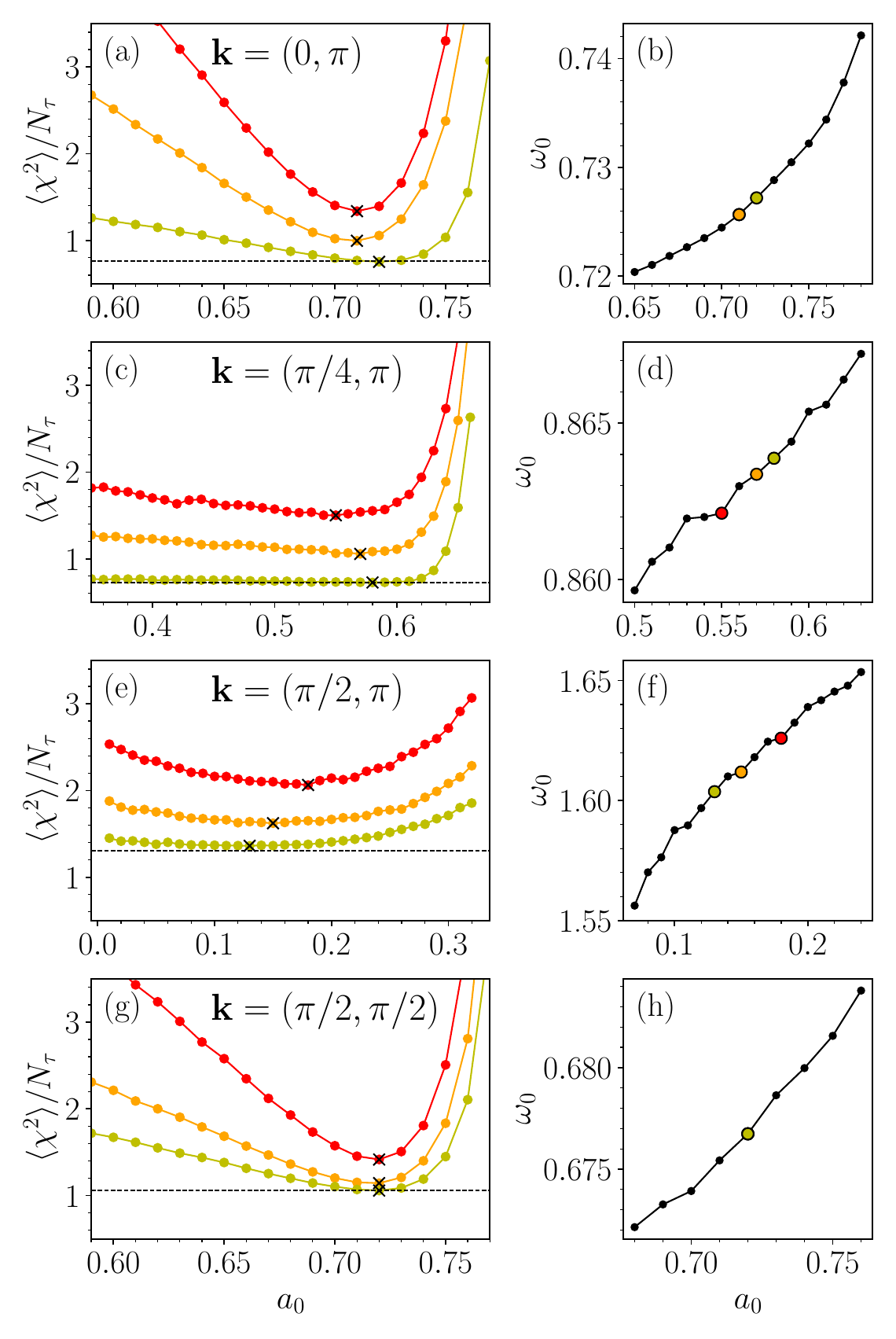}
\vskip-2mm
\caption{Scans over the $\delta$ weight $a_0$ in \Ak\ at four select momenta ($L = 16$, $U/t = 4$). The three colors correspond to three gradually lowered sampling temperatures (red to orange to yellow), where for the lowest $\Theta$ value, the $\chi^2$ minimum coincides with the simple criterion typically used to fix $\Theta$ when performing SAC (black dashed lines).}
\label{a0_scan}
\end{figure}

Here, we present exemplary results for the procedure used to determine the optimal value of the macroscopic $\delta$-function weight $a_0$ using a scan.
In Fig.~\ref{a0_scan} we show the scans used to determine $a_0$ for spectra between the $\Sigma$ and X points shown in Fig.~\ref{Ak}.
In the left column, $\langle \chi^2\rangle$ is plotted versus $a_0$, and in the right column the corresponding location of the edge ($\omega_0$) is shown. In the cases where the location of the $\langle \chi^2\rangle$ minima drifts as the $\Theta$ is lowered, the edge location (colored points in the right panels) shifts only moderately. The minimum flattens as $\Theta$ is lowered and becomes hard to discern below the value corresponding to the optimal sampling temperature, indicated with dashed lines. We note that the minimum is the sharpest for the largest $a_0$ values; for the $\mathbf{k}$-points closest to the non-interacting Fermi surface. These BZ points are also the most important ones for the purposes of this work, as will be discussed in the following section.

\subsection{Single-Particle Dispersion}

\begin{figure}[t]
\centering
\includegraphics[width=0.9\columnwidth]{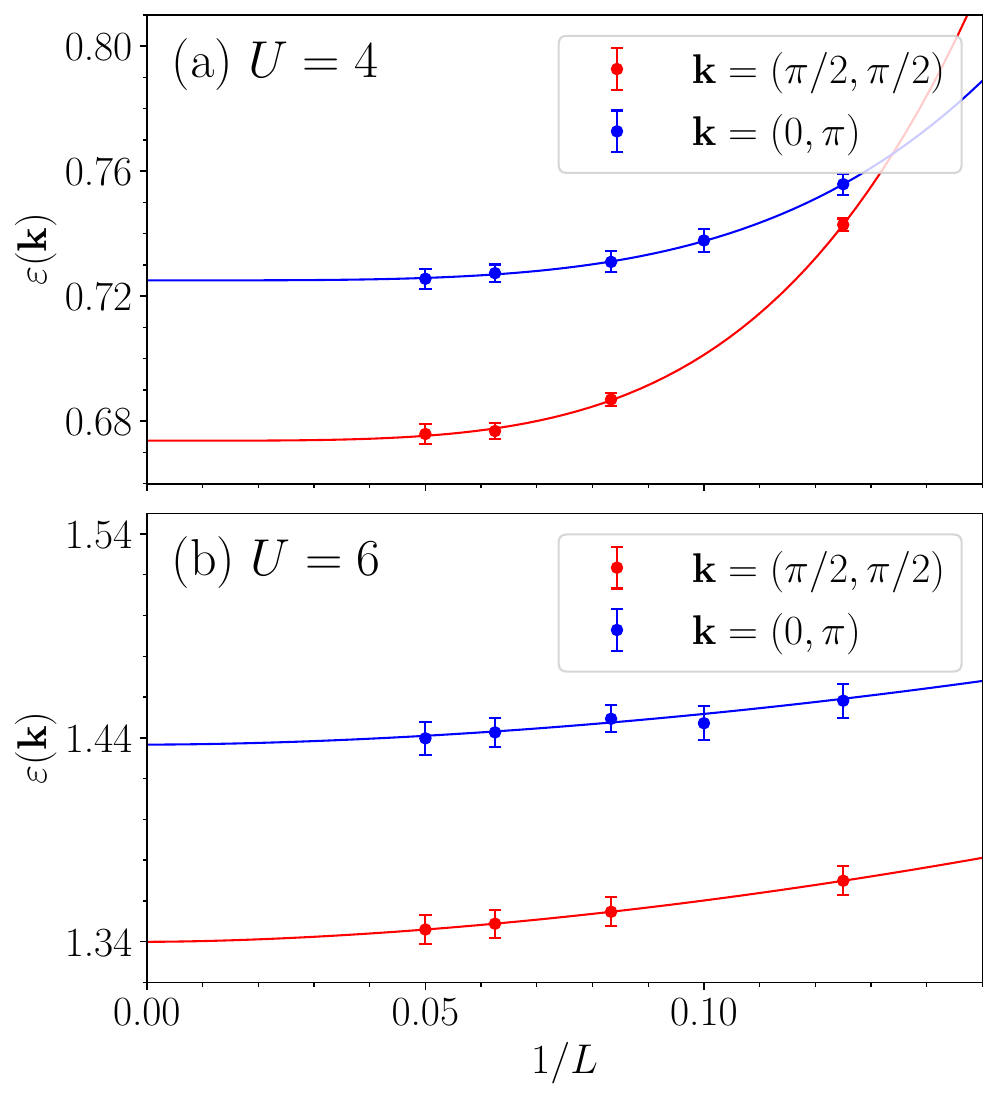}
\vskip-2mm
\caption[$\omega_{\rm X}$ (blue) and $\omega_\Sigma$ (red) as a function of system size for $U/t=4,6$.]{$\omega_{\rm X}$ (blue) and $\omega_\Sigma$ (red) as a function of system size for $U/t=4$, panel (a), and $U/t=6$, panel (b). The solid lines are power law fits $\omega(L) = a + b L^{-c}$. Each energy converges with $L$, with the difference between the $L=16$ and $L=20$ values being $<$1\% for both $U/t$ values. We estimated the uncertainty in the energy values by monitoring how the location of the macroscopic $\delta$-function edge changes as $a_0$ is slightly increased and decreased from its optimal value, as depicted in the right column of Fig.~\ref{a0_scan}}
\label{fss}
\end{figure}

The dispersion relation corresponding to the $\delta$-edge locations is shown along three lines in the BZ in Fig.~\ref{Ak}(b).
Unlike the noninteracting system, the $\Sigma$ and X states are no longer degenerate, with the former being the lowest in energy.
We have carried out these calculations for many system sizes for both $U/t=4$ and $U/t=6$ and show the size dependence of
$\omega_\Sigma$ and $\omega_{\rm X}$ in Fig.~\ref{fss}. There is no sign of the energy difference vanishing as $L \to \infty$, which was
suggested previously \cite{assaad_99}. The extrapolated difference $\omega_{\rm X}-\omega_\Sigma$ is about $0.05$ for $U/t=4$ and $0.1$ for $U/t=6$.

\begin{figure}[t]
\centering
\includegraphics[width=0.9\columnwidth]{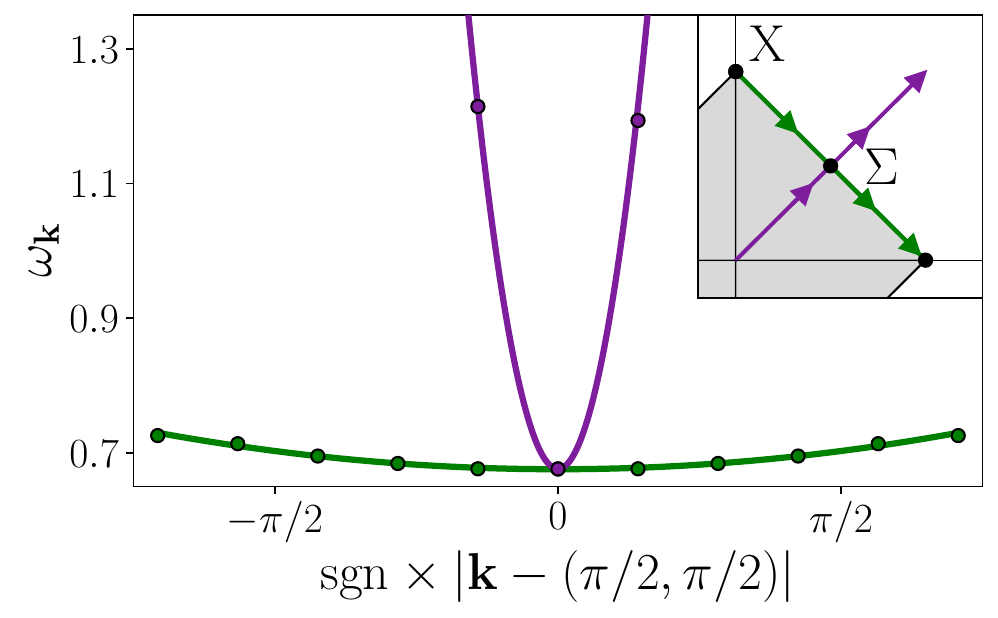}
\vskip-2mm
\caption{Quasiparticle dispersion along (green) and perpendicular to (purple) the non-interacting Fermi surface as a function of the signed distance in $\mathbf{k}$-space to $\Sigma$ (for $L=20$ and $U/t=4$). Just as in  Fig.~\ref{Ak}(b), the sign on the $x$-axis corresponding to the direction of the arrows in the right inset. The green and purple curves are both quadratic fits to the data, with the green fit the same as that shown in Fig.~\ref{Ak}(b).}
\label{quadratic_perp}
\end{figure}

We next consider the functional form of the dispersion about the $\Sigma$ and X points. It was previously argued that the dispersion is quartic
around X, supporting a metal-insulator transition with dynamic exponent $z=4$ \cite{assaad_96, imada_98, assaad_98, assaad_99, misawa_07}. Our
data can also be very well fitted to a quartic form, as shown in Fig.~\ref{Ak}(b), except on the line connecting $\Sigma$ and X (i.e., along the noninteracting
Fermi surface). Examining lines extending from the X point at other angles, we find that the energy drops below $\omega_{\rm X}$ only along the noninteracting
Fermi surface, while elsewhere the dispersion is approximately quartic above $\omega_{\rm X}$. A very small maximum in the energy exactly at the X point can 
be discerned in Fig.~\ref{Ak}, which might be taken as just an insignificant anomaly related to the resolution of the method. However, such a maximum has
also been systematically observed in the $t$-$J$ model, where it is small in the $J/t$ range corresponding to the $U$ values studied here but becomes very prominent
for larger $J/t$ \cite{brunner_00, mishchenko_01, lavalle_01,mishchenko_06}. Thus, we believe that the very small maximum also exists in the Hubbard model. In the
absence of the maximum, for a perfect quartic dispersion around the X points, the density of states originating from the neighborhood of the X points is divergent,
of the form $D(\omega \to \omega_{\rm X}^+) \propto (\omega - \omega_{\rm X})^{-1/2}$. The small deviations from the quartic form, the $X$ maximum in particular, will
quench the divergence but only in an insignificant way that we will disregard here. In the narrow range $\omega \in [\omega_\Sigma, \omega_{\rm X})$, the DOS
should be roughly constant, on account of the anisotropic quadratic, dispersion around the $\Sigma$ points; see Fig.~\ref{Ak}(b) and Fig.~\ref{quadratic_perp}.

\begin{figure}[t]
\includegraphics[width=0.97\columnwidth]{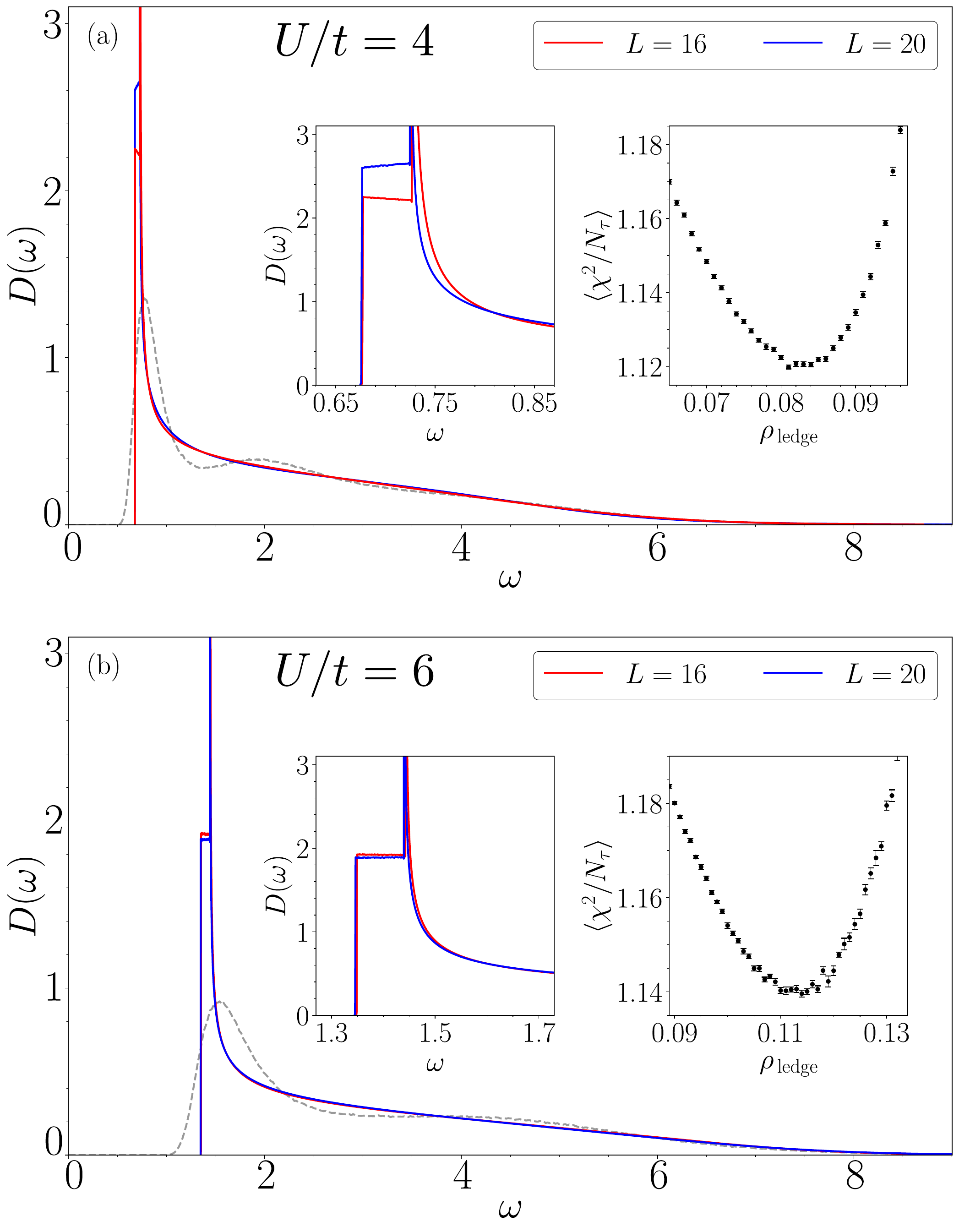}
\caption{DOS obtained with the double-edge SAC parameterization for both $L=16$ (blue) and $L=20$ (red), with $U/t=4$ in (a) and $U/t=6$ in (b).
The dashed grey lines are the results from unconstrained sampling for $L=20$. The left insets show zoomed in views of the ledge region
between $\omega_\Sigma$ and $\omega_{\rm X}$. The right insets show the goodness-of-fit versus the ledge fraction
$\rho_{\rm ledge}$ for $L=20$. The spectra shown were obtained using $\rho_{\rm ledge}$ of the minimum; in (a) $\rho_{\rm ledge}=0.071$ for $L=16$
and $\rho_{\rm ledge}=0.082$ for $L=20$, while in (b) $\rho_{\rm ledge}=0.115$ for $L=16$ and $\rho_{\rm ledge}=0.113$ for $L=20$.}
\label{dos}
\end{figure}

\section{Full Density of States}

Our results for the dispersion relation can now be fed in as prior information in SAC to extract the full DOS
from $G_{\rm loc}(\tau)$. We know the singular points $\omega_\Sigma$ and $\omega_{\rm X}$ and the smooth behavior that is expected between them, so we can use the parametrization in Fig.~\ref{params}(a) to resolve the DOS in this region. For $\omega \ge \omega_{\rm X}$, we use the constrained parametrization in Fig.~\ref{params}(c), with the lowest of the $\delta$-functions locked at $\omega_{\rm X}$. To determine the fraction $\rho_{\rm ledge}$
of states (i.e., the spectral weight) in the ledge of the DOS in $[\omega_\Sigma,\omega_{\rm X})$, a scan is performed over this parameter to locate a goodness-of-fit minimum, in analogy to the scan over $a_0$ of the $\delta$-edge in Fig.~\ref{a0_scan}. Results at the optimal value of  $\rho_{\rm ledge}$ are shown in Fig.~\ref{dos} along with the
  scan of $\langle \chi^2 \rangle$ versus $\rho_{\rm ledge}$. For reference, we also include results obtained by unconstrained sampling of the entire DOS, which cannot resolve
  the singularities and look very similar to previous results generating using MEM \cite{bulut_94,assaad_98}.

Based on the rapid size convergence of $\omega_{\Sigma}$, $\omega_{\rm X}$, and  the DOS (using the $L$-specific values of $\omega_{\Sigma}$ and
$\omega_{\rm X}$), we believe that the results in Fig.~\ref{dos} well represent the thermodynamic limit.
For both $U/t$ values, the DOS between the two singular points is indeed very flat, motivating our designation of this part as the ledge. The singular peak above $\omega_{\rm X}$ is followed by a thick tail with significant weight all the way up to $\omega = 6 \sim 7$.
Given that the $\chi^2$ values at the minimum of the scans are statistically good (with the sampling temperature at its optimal value
\cite{shao_23,schumm_24}), there is no statistical evidence for any additional peaks beyond the edge at $\omega_{\rm X}$. Such additional peaks, often referred to as ``ringing'', are common in MEM results and, as seen in Fig.~\ref{dos}, are also produced by unconstrained SAC. Ringing has been explained as a compensating behavior stemming from the presence of spectral weight inside a true gap \cite{shao_23}. The fact that this feature appears in both our unconstrained SAC results and the previous MEM results, but not in our gapped, constrained SAC results, strongly suggests that these peaks are indeed artifacts of an imperfect numerical analytic continuation procedure. By using our advanced, combined constrained SAC parameterization, we are able to resolve the detailed structure of the DOS with high precision, and produce a spectrum containing no spurious
ringing features.

\section{Implications for the Doped Hubbard Model}

Our results demonstrate that the smallest gap in the Hubbard model is clearly at $\Sigma$, and we confirm a dispersion close to quartic around X. There
is a barely discernable local X maximum, seen in Fig.~\ref{Ak}(b), which implies a minor rounding of the singular DOS at $\omega_{\rm X}$ that cannot
be resolved with our methods. Having established these facts, we determined the
fraction $\rho_{\rm ledge}$ of states below $\omega_{\rm X}$. Within the RBA, four hole pockets would form upon light doping $x>0$ (as observed in the form of ``Fermi arcs'' in underdoped cuprates \cite{tanaka_06,valla_06}), and merge into a contiguous Fermi sea as the energy approaches $\omega_{\rm X}$.

The doping at which ledge states will be exhausted and the Fermi surface will reconnect, $x_c$, would at first sight be be equal to $\rho_{\rm ledge}$, $\sim 0.1$ for the two $U/t$ values considered here. However, this estimate of $x_c$ neglects the fact that there is a significant continuum extending rather far above $\omega_X$, corresponding to a collection of excited states that dress the $\omega_{\mathbf{k}}$ quasiparticles. A simple way to correct for the fact that only the peak contribution to \Ak\ is accounted for in $\rho_{\rm ledge}$ is to divide by the quasiparticle weight $a_0$ for the states with $\omega_{\mathbf{k}}$ within the ledge, $\sim 0.7$ for $U/t=4$ and $\sim 0.6$ for $U/t = 6$. A more accurate approach to calculating $x_c$ in the RBA is by direct counting of the fraction $n_{\rm ledge}$ of quasiparticle energies below $\omega_{\rm X}$.

\begin{figure}[t]
\includegraphics[width=0.72\columnwidth]{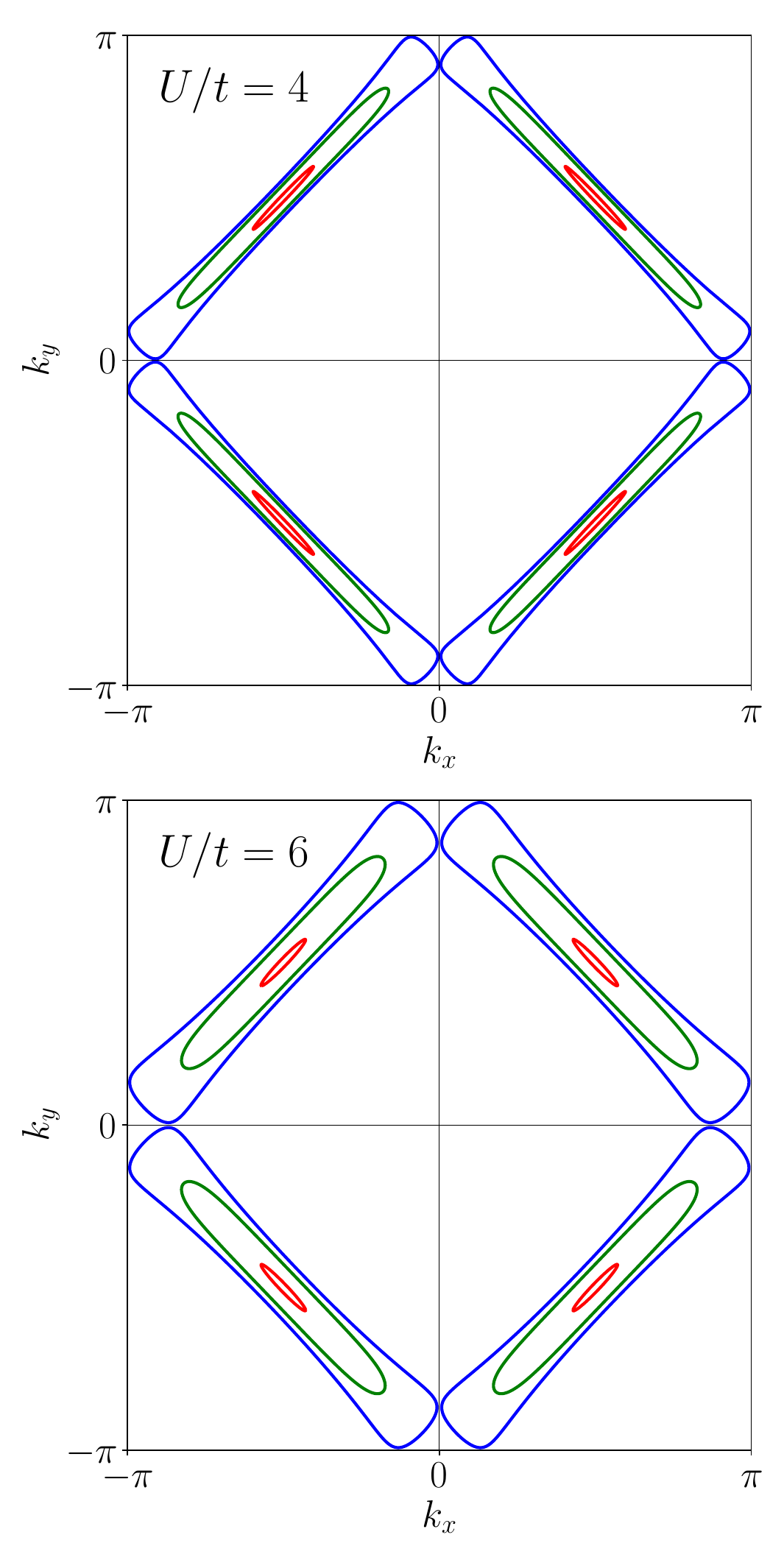}
\vskip-2mm
\caption{Fermi surfaces at doping levels corresponding to three energy levels: just above $\omega_\Sigma$ (red), just below $\omega_{\rm X}$ (blue), and an intermediate energy (green).}
\label{fermi_surface}
\end{figure}

To do this, we performed a series of systematic quadratic fits to the dispersion across the noninteracting Fermi surface (e.g. purple in Fig.~\ref{quadratic_perp}). The perpendicular quadratic dispersion gradually and continuously flattens as $\mathbf{k}$ approaches X, and the quadratic coefficient can be well fit to a cosine function, though this form is likely only approximate but sufficiently precise for our purposes. Using this function, along with the quadratic fit along the noninteracting Fermi surface (green in Fig.~\ref{quadratic_perp}), we can model the quasiparticle dispersion in the vicinity of $\Sigma$ by a continuous function $\omega(\mathbf{k})$. This allows us to map out the Fermi surface at any level of doping in the RBA by identifying equal energy contours, as shown in Fig.~\ref{fermi_surface} for both $U/t=4$ and $U/t=6$. We note that as the energy approaches $\omega_{\rm X}$ (the blue contour), the Fermi surface flattens near the X point, reflecting that  $\omega_{\rm X}$ corresponds to a slight local maximum, as discussed above. 

The $\mathbf{k}$-space area of the contour with energy $\omega_{\rm X}$ thus provides an alternative estimate for the critical level of doping $x_c$ under the RBA. Our calculations give critical doping levels of  $x_c= n_{\rm ledge} \approx 0.13$ and $0.20$ for $U/t=4$ and $6$, respectively. Both $x_c$ values are larger than the values of $\rho_{\rm ledge}$ extracted from the DOS by nearly exactly a factor of $1/a_0$, just as predicted. Verifying that this alternative approach yields a value for $x_c$ consistent with out prediction $n_{\rm ledge} \approx \rho_{\rm ledge} / a_0$ servers as a strong consistency check for our methodology and our application of the RBA in this context.

\section{Conclusion and Discussion}

Despite the fundamental importance of the 2D Hubbard, crucial details
of the quasiparticle dispersion relation and the density of states have been lacking. It has for long been established that the related $t$-$J$ model
hosts its lowest $x=0$ quasiparticle for small $J/t$ at ${\bf k} = \Sigma$, and a rather flat band around ${\bf k} = {\rm X}$
has been observed (whereas a very significant local maximum appears at X for larger $J/t$) \cite{brunner_00, mishchenko_01, lavalle_01,mishchenko_06}.
The energy splitting of the $\Sigma$ and X
quasiparticles is also a well documented feature of the underdoped cuprates \cite{opel_00,letacon_06,tanaka_06,valla_06,millis_06,aichhorn_07,hufner_08}.
However, this behavior has not been established in the case of the Hubbard model at moderate $U$ values which are relevant to the cuprates
\cite{assaad_98,assaad_99,imada_98}.

Our results also have intriguing implications for the behavior of the 2D Hubbard model away from half-filling, particular in connection to the cuprate superconductors. The exact values of $n_{\rm ledge}$ obtained here should not be taken as specific predictions for the cuprates, as there will
clearly be significant effects of interactions beyond the Hubbard model. Indeed, the rather large dependence on $U$ can be taken as a general
high sensitivity of $n_{\rm ledge}$ to model parameters. Our main point here is the presence of the second singularity at $\omega_X$, which should imply a
drastic change in the doped state at $x_c=n_{\rm ledge}$, a change from a ``plain'' doped Mott insulator \cite{lee_06, proust_19}, likely with strong
spin and charge density correlations \cite{xu_24}, to something else.
While it appears plausible that the RBA applies at low doping in the Hubbard model, in some
calculations, a CDW or stripe instability takes place that would likely have to involve breakdown of
the rigid band and a Fermi surface reconstruction \cite{taillefer_10, freire_15, badoux_16, proust_19}. Furthermore, it is now believed that Luttinger's theorem is violated in the underdoped cuprates \cite{yoshida_03, senthil_03, mei_12} due to the emergence of topologically ordered states, as investigated
within a fermion-doped dimer-singlet model \cite{kaul_07, sachdev_16a, sachdev_16b}. If this pictures applies immediately upon doping, it would invalidate the application of the RBA to describe the low doping behavior of these systems.  But in the absence of such complications (which may rely on a next-nearest neighbor hopping $t'$ to frustrate
the antiferromagnetic ordering tendencies close to half-filling), a compelling scenario emerging from our study of the standard Hubbard model is the rigid band breaking
down only when the doping exceeds the ledge fraction at $x_c = n_{\rm ledge}$. A reason for this doping induced Fermi surface reconstruction could be the $(\pi,\pi)$
scattering (magnon exchange) between the X points with their high density of states.

Though the standard Hubbard model may not itself have a superconducting phase \cite{qin_20, qin_22}, the second singularity in the DOS could still correspond to a
critical point. The extended superconducting phase would then be induced only in the presence of additional interactions, with $t'$ hopping the most promising candidate so far \cite{hirayama_18,xu_24}.

The mechanism of high-temperature superconductivity is still an open question. The Hubbard model remains a key piece of this puzzle and our
presented results provide further compelling insights for exploring their deep connection. The sharp spectral features that we have resolved here are
currently challenging to observe directly in experiments, because of limitations of  ARPES \cite{shen_04, valla_12, drachuck_14} and scanning-tunnelinng
miscroscopy \cite{davis_15} in the half-filled, Mott-insulating regime, but our results will
hopefully stimulate novel experimental tests.

\begin{acknowledgments}
We would like to thank Fahker Assaad, Masatoshi Imada, Mohit Randeria, Subir Sachdev, Nandini Trivedi, and Ettore Vitali for stimulating discussions,
and Chunhan Feng and Yuan-Yao He for assistance with computation and software. This research was supported by the Simons Foundation
under Grant No. 511064. Computational resources were provided by the Shared Computing Cluster managed by Boston University’s Research Computing Services
and by the Flatiron Institute Scientific Computing Center. The Flatiron Institute is a division of the Simons Foundation.
\end{acknowledgments}

\bibliography{bib.bib}
\end{document}